\documentclass[
    ,final            
  ]
  {aipproc}

\layoutstyle{6x9}

\begin{document}

\title{Probing Supersymmetry with Neutral Current Scattering Experiments}

\author{A. Kurylov}{
  address={California Institute of Technology,
Pasadena, CA 91125\ USA}
}

\author{M.J. Ramsey-Musolf}{
  address={California Institute of Technology,
Pasadena, CA 91125\ USA},
altaddress={Department of Physics, University of Connecticut,
Storrs, CT 06269\ USA}
}

\author{S.Su}{
  address={California Institute of Technology,
Pasadena, CA 91125\ USA}
}

\begin{abstract}
We compute the supersymmetric contributions to
the weak charges of the electron ($Q_W^e$) and proton ($Q_W^p$)
in the framework of Minimal
Supersymmetric Standard Model.  
We also consider the ratio of neutral
current to charged current cross sections, $R_\nu$ and $R_{\bar\nu}$
at $\nu$ (${\bar\nu}$)-nucleus deep inelastic scattering,
and compare the supersymmetric corrections with the deviations
of these quantities from the Standard Model predictions implied by the
recent NuTeV measurement.
\end{abstract}

\maketitle


\section{Introduction}
\label{sec:intro}

In the Standard Model (SM) of particle physics, 
the predicted  running of $\sin^2\theta_W$ from $Z$-pole
to low energy: $\sin^2\theta_W(0)-\sin^2\theta_W(M_Z)=0.007$, 
has never been established experimentally to a high
precision.  $\sin^2\theta_W(M_Z)$ can be obtained through
the $Z$-pole precision measurements with very small error.  However, 
no determination of $\sin^2\theta_W$ at low energy with similar precision
is available.  More recently, the results of cesium atomic
parity-violation (APV) \cite{Ben99} and $\nu$- ($\bar\nu$-)
nucleus deep inelastic scattering (DIS)\cite{NuTeV}  
have been interpreted as determinations of the
scale-dependence of $\sin^2\theta_W$. The cesium APV result appears to be
consistent with the SM prediction for $q^2\approx 0$, whereas the neutrino
DIS measurement implies a $+3\sigma$ deviation at $|q^2|\approx 100$
${\rm GeV}^2$. If
conventional hadron structure effects are ultimately unable to
account for the NuTeV ``anomaly", the results of this precision
measurement would point to new physics.

In light of this situation, two new measurements involving polarized
electron scattering
have taken on added interest: parity-violating (PV) 
M\"oller ($ee$) scattering at
SLAC\cite{slac} and elastic,  PV
$ep$ scattering at the Jefferson Lab (JLab)\cite{qweak}. In the absence of
new physics, both
measurements could be used to determine $\sin^2\theta_W$ at the same scale:
$|q^2|\approx 0.03$ ${\rm GeV}^2$, with comparable precision in
each case: $\delta \sin^2\theta_W =0.0007$.  
Furthermore, the precision needed to
probe new physics effects, e.g. supersymmetry (SUSY),
is roughly an order of magnitude less stringent,
owing to a fortuitous suppression of the SM electron and proton
weak charge: $Q_W^p=-Q_W^e=1-4\sin^2\theta_W\approx 0.1$ 
at tree-level.  Consequently,
experimental precision of order a few percent, rather than a
few tenths of a percent, is needed to probe new physics corrections.

The goal of our study is to develop consistency check for theories 
of new physics using the low energy neutral current scattering measurements.  
In particular,
we will consider the Minimal Supersymmetric Extension of 
SM (MSSM)\cite{haber-kane}, which is the most 
promising candidate for new physics beyond SM.  
For $R$-parity conserved MSSM, low-energy precision observables 
experience SUSY only via loop effects involving
virtual supersymmetric particles.  Tree level corrections 
appear once $R$-parity is broken explicitly.  
We studies both the PV electron scattering (PVES) and 
$\nu$ ($\bar \nu$)-nucleus DIS processes.
Details of the calculations presented here
can be found in Ref.~\cite{apv} and \cite{kur-rm-su-nutev}.

\section{Radiative Contribution to Weak Charge}
\label{sec:formalism}

The weak charge of a particle $f$ is defined as the strength
of the effective $A(e)\times V(f)$ interaction:
${\cal L}_{EFF}^{ef}=-{G_\mu\over 2\sqrt 2}Q_W^f {\bar e} \gamma_\mu\gamma_5
e {\bar f}\gamma_\mu f
$.
With higher-order corrections included, the weak charge 
can be written as
$Q_W^f = \rho_{PV}\left[2 T_3^f 
-4 Q_f\kappa_{PV}\sin^2\theta_W\right]+\lambda_f
$.
The quantities
$\rho_{PV}$ and $\kappa_{PV}$ are universal, while 
the correction $\lambda_f$, on the other hand, does depend on the
fermion species.  At tree-level, one has $\rho_{PV}=1=\kappa_{PV}$ and
$\lambda_f=0$, while at one-loop order
$\rho_{PV}=1+\delta \rho^{\rm SM}_{PV} + \delta \rho^{\rm SUSY}_{PV}$, 
and similar 
formulae apply to $\kappa_{PV}$ and $\lambda_f$.

The counterterm $\delta {\hat G_\mu}$ determined by muon
life time and the $Z^0$ boson self-energy 
are combined into $\rho_{PV}$
(expressed in terms of the oblique parameters $S$, $T$ \cite{stu-degrassi}):
\begin{equation}
\rho_{PV}=1+{{\delta {\hat G_\mu}} \over G_\mu}+{{\hat \Pi}_{ZZ}(0)\over M_Z^2}
=1-{{\hat \Pi}_{WW}(0)\over M_W^2}+{{\hat \Pi}_{ZZ}(0)\over M_Z^2}-{\hat
\delta}_{VB}^\mu=1+ {\hat\alpha} T-{\hat\delta}_{VB}^\mu.
\end{equation}
The quantity ${\hat \delta}_{VB}^\mu$ denotes the
the electroweak vertex, external leg, and box graph corrections to the muon
decay amplitude.

$Z-\gamma$ mixing and parity-violating electron-photon
coupling $F_A^e(0)$
contribute to $\kappa_{PV}$:
\begin{eqnarray}
\label{eq:kappa}
\kappa_{PV}&=&1+{{\hat c}\over {\hat s}}
{{\hat \Pi}_{\gamma Z}(q^2)\over q^2}+4{\hat c}^2 F_A^e(0)+{\delta{\hat
s}^2_{\rm new}\over {\hat s}^2}
=1+\left({{\hat c}^2\over {\hat c}^2-{\hat s}^2} \right)
\left({{\hat\alpha}\over 4{\hat s}^2 {\hat c}^2} S-{\hat \alpha} T
+{\hat\delta}_{VB}^\mu \right)
 \nonumber \\
&&
\hspace{-0.5 in}
+{{\hat c}\over {\hat s}}\Bigl[  {{\hat\Pi}_{Z\gamma}(q^2)\over q^2}-
{{\hat\Pi}_{Z\gamma}(M_Z^2)\over M_Z^2}\Bigr]
+\Bigl({{\hat c}^2\over {\hat c}^2-{\hat s}^2}
\Bigr)\Bigl[-{{\hat\Pi}_{\gamma\gamma}(M_Z^2)\over M_Z^2}
+{\Delta{\hat\alpha}\over \alpha}
\Bigr]+4{\hat c}^2 F_A^{e}(q^2)
\end{eqnarray}
The shift ${\delta{\hat s}^2_{\rm new}}$ in ${\hat s}^2$
follows from the definition of ${\hat s}^2$ 
in terms of $\alpha$, $G_\mu$, and $M_Z$\cite{apv}.

The non-universal contribution $\lambda_f$ 
to the weak charge is determined by the sum
of the the renormalized vertex
corrections and the box graphs.

\section{SUSY Correction to Weak Charges}
\label{sec:weakcharge}
\begin{figure}
\resizebox{5.8cm}{!}{
\includegraphics{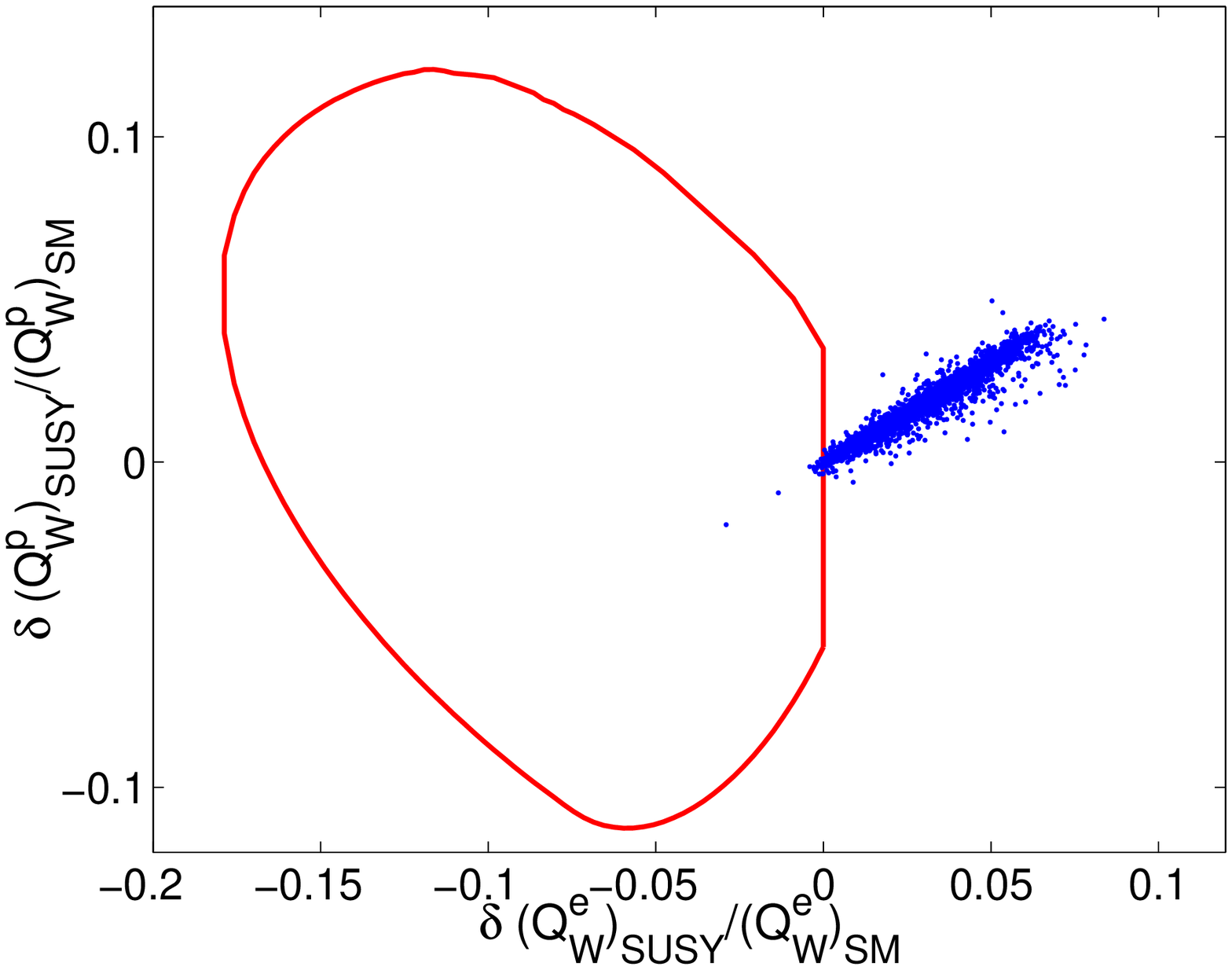}}
\resizebox{5.8cm}{!}{
\includegraphics{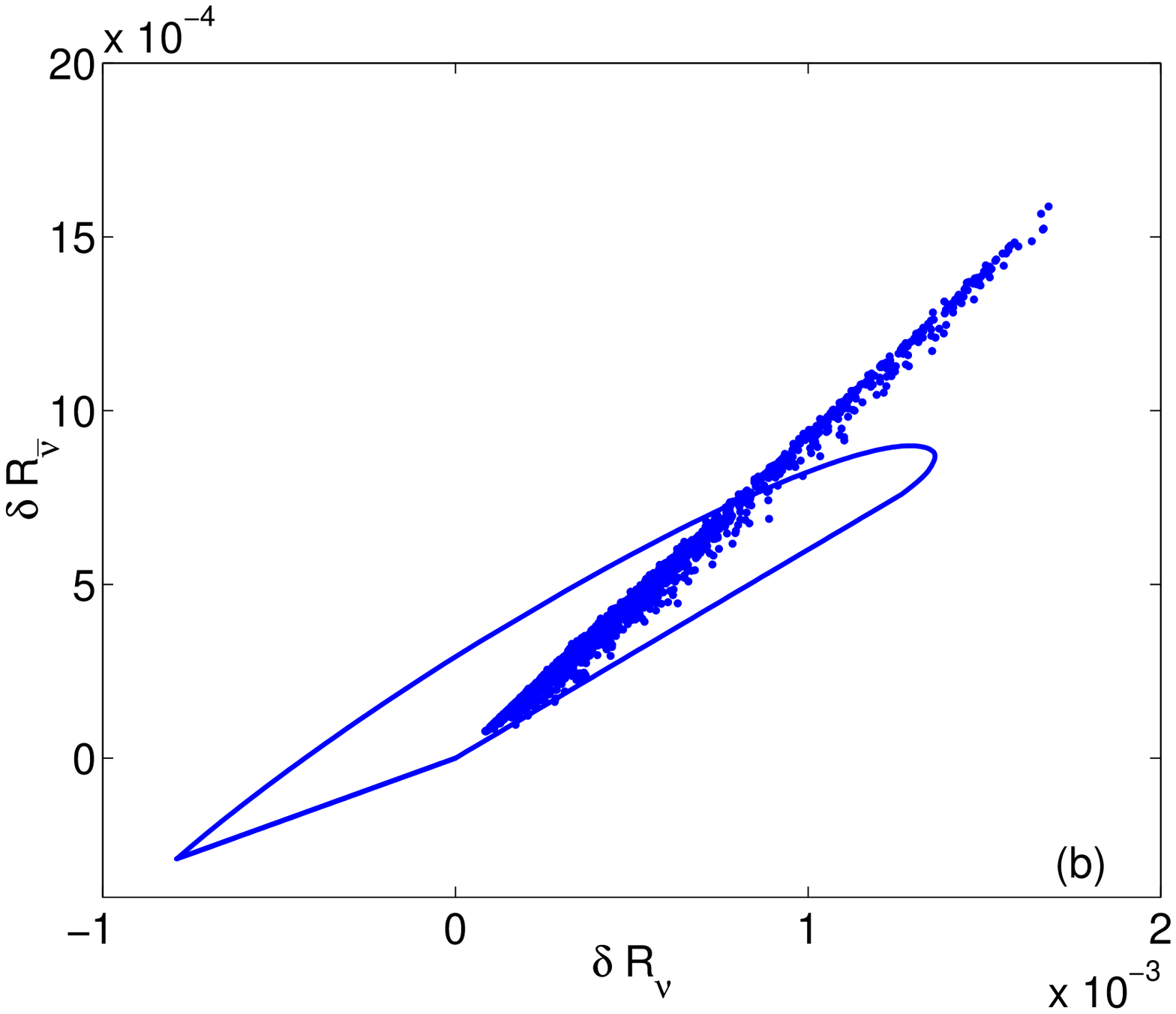}}
\caption{Plot(a) shows the relative shifts in electron and proton weak charges
due to SUSY effects. 
Plot(b) shows the MSSM contribution to $R_{\nu}$ and $R_{\bar \nu}$.
Dots indicate MSSM loop corrections for $\sim 3000$
randomly-generated SUSY-breaking parameters. Interior of truncated
elliptical region gives possible shifts
due to RPV SUSY interactions (95\% confidence).}
\label{fig:apv_NuTeV}
\end{figure}
In order to evaluate the potential size of SUSY loop corrections,
a set of about 3000 different combinations of SUSY-breaking
parameters was generated.
Fig.~\ref{fig:apv_NuTeV}(a) shows the shift in the weak charge
of the proton, $\delta Q_W^p = 2\delta Q_W^u+ \delta Q_W^d$, versus
the corresponding shift in the electron's weak
charge, $\delta Q_W^e$, normalized to the respective SM values.
The corrections in the
MSSM (with
$R$-parity conserved) can be as large as
$\sim 4\%$ ($Q_W^p$) and
$\sim 8\%$ ($Q_W^e$) -- roughly the size of the proposed experimental
errors for the two PVES measurements.
The shifts $\delta Q_W^{e,p}$ are dominated by $\delta\kappa^{\rm
SUSY}_{PV}$, which is nearly always negative, corresponding to
a reduction in the  value of
$\sin^2{\theta}_W^{eff}(q^2)=\kappa_{PV}(q^2)\sin^2{\theta}_W$
for the PVES experiments.
Since this effect is identical for both
$Q_W^e$ and $Q_W^p$, the dominant effect of
$\delta \kappa_{PV}$ 
produces a linear correlation between the two weak charges.

As evident from Fig.~\ref{fig:apv_NuTeV} (a), the relative sign of
the loop corrections to both $Q_W^p$ and
$Q_W^e$ is nearly
always the same and positive. 
This correlation is significant, since the effects of other new
physics scenarios can display different signatures. For example, for the
general class of theories based on $E_6$ gauge group, with neutral gauge
bosons $Z^{\prime}$ having mass $< 1000$ GeV, the effects on
$Q_W^p$ and $Q_W^e$ also correlate, but $\delta Q_W^{e,p}/Q_W^{e,p}$ can
have either sign in this
case\cite{MRM99,Erl-MJRM-Kur02}. 
In contrast, leptoquark interactions would not lead
to discernible
effects in $Q_W^e$ but could induce sizable shifts in 
$Q_W^p$ \cite{MRM99,Erl-MJRM-Kur02}.

As a corollary, we also find that SUSY loop
corrections to
the weak charge of cesium is suppressed: 
$\delta Q_W^{\rm Cs}/Q_W^{\rm Cs}< 0.2\%$ 
and is equally likely to have either sign, 
which is smaller than the  presently quoted
uncertainty for the cesium nuclear weak charge of
about 0.6\% \cite{sushov02}.
Therefore, the present agreement
of $Q_W^{\rm Cs}$ with the SM prediction does not preclude significant shifts
in $Q_W^{e,p}$ arising from SUSY.
The situation is rather different, for example, in the $E_6~Z^\prime$
scenario, where sizable shifts in $Q_W^{e,p}$ would also imply observable
deviations of $Q_W^{\rm Cs}$ from the SM prediction.

New tree-level SUSY contributions to the weak charges can be generated
when the R parity in MSSM is not
conserved.  The effects of $R$-parity violating (RPV) 
contribution can be parametrized by 
positive,
semi-definite, dimensionless quantities
$\Delta_{ijk}({\tilde f})$ and $\Delta_{ijk}^\prime({\tilde f})$\cite{MRM00}, 
which are constrained from the existing precision
data \cite{MRM00}. 
The 95\% CL region allowed in the $\delta Q_W^p/Q_W^p$ 
vs. $\delta Q_W^e/Q_W^e$ plane is
shown by the closed curve in Fig.~\ref{fig:apv_NuTeV} (a). 
We observe that the prospective effects of
RPV are quite distinct from SUSY loops. 
The value of $\delta Q_W^e/Q_W^e$ is
never positive in contrast to the situation for SUSY loop effects, 
whereas $\delta Q_W^p/Q_W^p$
can have either sign.

Thus, a comparison of the two PVES measurements could help
determine which extension of the MSSM is to be favored over other new
physics scenarios \cite{Erl-MJRM-Kur02}.

\section{NuTeV Measurement}
\label{sec:nutev}

Recently, the NuTeV collaboration has performed a
precise determination of the ratio $R_\nu$
($R_{\bar \nu}$) of neutral and 
charged current deep-inelastic
$\nu_\mu$ ($\bar\nu_\mu$)-nucleus cross sections\cite{NuTeV},
which can be expressed in terms of the effective $\nu-q$ hadronic couplings
$(g_{L,R}^{\rm eff})^2$:
\begin{equation}
\label{eq:rnudef}
R_{\nu ({\bar\nu})} = 
\frac{\sigma(\nu (\bar \nu) N \rightarrow \nu  (\bar \nu) X)}
{\sigma (\nu  (\bar \nu) N\rightarrow l^{-(+)} X)}
= (g_L^{\rm eff})^2 + r^{(-1)} (g_R^{\rm eff})^2\ \ \ ,
\end{equation}
where $r=\sigma^{CC}_{{\bar\nu} N}/\sigma^{CC}_{\nu N}$.
Comparing
the SM predictions\cite{pdg} for $(g_{L,R}^{\rm eff})^2$ with the values
obtained by the NuTeV
Collaboration yields deviations
$\delta R_{\nu({\bar\nu})}=
R_{\nu({\bar\nu})}^{\rm exp}-R_{\nu({\bar\nu})}^{\rm SM}$, 
$\delta R_{\nu}=-0.0029 \pm 0.0015$,
$\delta R_{\bar{\nu}}=-0.0015 \pm 0.0026$.

The numerical results for SUSY contributions 
to $R_\nu$ and $R_{\bar \nu}$ via the correction to the effective hadronic 
couplings $(g_{L,R}^{\rm eff})^2$
are  shown in Fig.~\ref{fig:apv_NuTeV} (b).
For detailed analysis, see 
Ref.~\cite{kur-rm-su-nutev}.
SUSY loop contributions to $R_\nu$ and $R_{\bar \nu}$ 
are smaller
than the observed deviations.
More significantly, the sign of the SUSY loop
corrections is
nearly always positive, in contrast to the sign of the NuTeV anomaly.
Tree-level RPV contributions to 
$R_\nu$ and $R_{\bar \nu}$ are
by and large positive. While small negative corrections are also possible,
they are numerically
too small to be interesting.

\section{Conclusion}
\label{sec:conclusion}
In summary, we have studied the SUSY corrections to the weak 
charge of the electron and proton, which could be measured at
PV $ee$ and $ep$ scattering experiments.  The correlation between
these two quantities could be used to distinguish various new
physics.  We also examined the SUSY contributions to the NuTeV
measurements and found that it is hard to explain the NuTeV 
anomaly in the framework of MSSM.

\end{document}